# Body mass index – a physics perspective

S. P. Apell, O. Wahlsten and H. Gawlitza
Department of Applied Physics, Chalmers University of Technology
SE-412 96 Göteborg, Sweden

**Abstract:**
Almost two centuries ago Adolphe Quetelet came up with an index to characterize man which is frequently used today to make predictions about health status. We show that this so called *body mass index* is directly related to the ratio between the physical quantities metabolic rate and heat loss.

## 1. A historical index

*Physics* from ancient Greek *physis* "nature" and *physiology* with the added ending *logia* "study of" have more than an etymological connection when it comes to *body mass index* as we will show in this communication.

Health status such as obesity is generally correlated with a range of cardiovascular, chronic and cancerous diseases as well as early death [1]. For this reason there is a medical interest in a reliable, cheap and easy way to monitor health risks by turning physiological features into a number. Such a number could also facilitate comparisons of health status between different regions, populations, ages and gender and for the individual when monitoring its temporal development. We will show that there is an obvious physical justification for such a powerful index.

Body mass index is not the only index or measurement of general health status. There are several alternative and complementary indices such as Ponderal Index, Body Volume Index, Skin Fold Method, Waist-to-Hip Ratio and Sagittal Abdominal Diameter [1]. Through a comparative study of indices of obesity A Keys in 1972 introduced the notion *body mass index* [2] for the best performing index. This was a renaming of the *Quetelet index* proposed in 1832 by the Belgian mathematician, astronomer and statistician Adolphe Quetelet. For a person of weight *W* and height *H* it is simply [3,4]:

$$BMI \equiv W/H^2, \qquad (1)$$

and has been the target of numerous scientific studies and is widely used by doctors and laymen alike. The Quetelet index was however not initially intended to be used to characterize obesity or general health status. It would rather help Quetelet define a "normal man" by fitting a Gaussian curve to the distribution he found for the index since using mass alone did not work [5].

The motivation for our study is to ask what can be the physical basis for the *body mass index*. We thus analyze it from a physical perspective with emphasis on dimensional analysis, scaling with mass and relevant length-scales such as waist circumference as well as relating it to body metabolic rate and heat loss. We limit our analysis to adults and refer the reader to [6] for aspects related to children.

## 2. Physical aspects of body mass index

Quetelet constructed the index based on his findings that the weight of adult humans scales with their height squared. If we were growing equally in all directions our weight would grow as the cube of our height, often used in many physics papers on animal scaling (see e.g. [7]), but this is the case only during our first year of growth. In this context we recall the well-known story Gulliver's Travels by Jonathan Swift, published in 1726. Both the Lilliputians and the Brobdingnagians were of the same proportions as Gulliver, but the Lilliputians were about 10 times shorter and the Brobdingnagians were about 10 times taller. This would result in the Lilliputians having a *body mass index* a factor 10 smaller than Gulliver and the Brobdingnagians a *body mass index* 10 times larger than Gulliver assuming the traditional scaling of body mass with the third power of the linear dimensions. However we know from Quetelet's observations that this is not the case.

Moreover Quetelet pointed at the height as an important variable in characterizing human populations apart from weight. The relevance of height is interesting from the physical point of view that we defy gravity by growing mainly in the opposite direction to the gravity field, probably related to some early evolutionary advantages in the development of man. We now look at the index from the point of view of dimensional analysis.

### 2.1 Dimensional analysis

Multiplying the *body mass index* with the gravitational acceleration it can be viewed as a pressure. Climbing our own height *H* the index can be interpreted in terms of an energy density. In both cases the numbers come out to be very small compared to other relevant physical pressures and energy densities such as air pressure and food intake per unit volume. The main reason is of course that it is the height which appears in the *body mass index*. H is at least one order of magnitude larger than the typical length scale of our contact area with the ground and this points us in the direction of relating the *body mass index* to a length-scale.

We therefore divide the *body mass index* with an average body density to relate it entirely to body geometrical properties in an obvious notation:

$$\frac{BMI}{\rho_{av}} = \frac{V}{H^2} \equiv \frac{<A>}{H} \equiv L_{mp} \qquad (2)$$

thus converting *body mass index* to a length-scale $L_{mp}$. Since height is almost constant for adults we see that *body mass index* is directly related to the average cross-sectional area defined by $<A> \equiv V/H$. The density normalized index thus becomes area over length, which we recognize as the typical scaling of a capacitor in electromagnetism: the density normalized index grows with the average cross-sectional area of the body and decreases with increasing height. Instead focusing on $L_{mp}$, Equation (2) shows that the *body mass index* can be seen as the thickness in millimeters of the meat plate created if you were to flatten yourself out into a square with side length equal to your height. This means for typical index values of 20-30 the corresponding thickness $L_{mp}$ ~ 20-30 mm. From energy considerations our height H compared to $L_{mp}$ is a measure of the gravitational prize for being erect.

## 2.2 Shape, weight and waist

For adults the height is an almost constant number throughout life and the variation in *body mass index* is solely related to changes in mass and thus changes in the average cross-sectional area *<A>* of the body. In other words one could forsee that different body shapes could actually have the same average area and *body mass index*.

Simplifications are always of importance when analysing physical aspects of complex objects and geometries. It would be natural for a physicist to represent a human with simple geometrical objects of rotational symmetry. Going beyond the spherical man approximation the simplest of them only need two parameters to specify; e.g. a cylinder or a prolate spheroid. The cylinder is of course extreme since it implies a human with a constant width and the prolate spheroid one with small brain and feet. Given the height and weight (i.e. volume) of a person, and hence a given *body mass index*, completely specifices these shapes. This illustrates how the same *body mass index* can correspond to different shapes (mass distributions), albeit they are very simplified. The prolate spheroid has an average cross-sectional area which is only 2/3 times that of the cylinder given they have the same height. In general terms we expect that the average cross-sectional area has the following form:

$$<A> = L_1 L_2 f(shape) \qquad (3)$$

where the *L's* are some relevant length-scales in the problem and *f(shape)*, expected to be of order unity, is a general function of the shape of the object. In Eq. (2) $L_1 = H$, $L_2 = L_{mp}$ and

*f=1*. For the cylinder case we can write $L_1 = L_2 = B$ (width) and *f(shape)* = π/4. The density normalized *body mass index* then becomes

$$\frac{BMI}{\rho_{av}} = \frac{\pi B^2}{4H} = \frac{\pi B}{4\eta} \propto C/\eta \tag{4}$$

where $\eta \equiv H/B$ is the cylinder aspect ratio and C the waist circumference. The same scaling goes for a prolate spheroidal modelling. We notice that this result is rather general for all bodies of rotational symmetry and only characterized by two length-scales, i.e. a scaling with inverse aspect ratio and waist circumference C. One of Quetelet's most important observations was however that $<A> = H$ [3]. A living body grows in such a way that the three length scales for "height, width and depth" are interrelated. In this sense there has to be yet another length-scale ($L_{mp}$) in the human growth problem to obtain the proper dimensions of $<A>$. We can at this stage only speculate that it is tied to a characteristic length scale for the fat deposits.

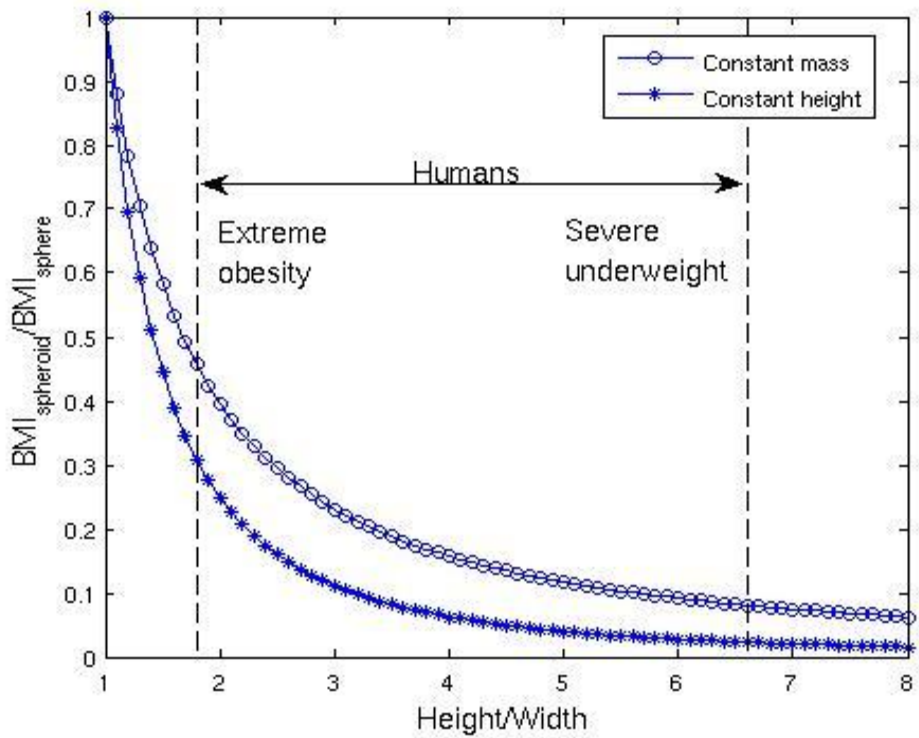

*Figure 1*: *Body mass index for a prolate spheroid in units of the body mass index for a sphere of the same mass (circles) or the same height (stars). Upper curve is for a population of individuals of the same mass and different heights and the lower curve is for a single individual with constant height but changing mass. In both cases we have a considerable variation over the range of human shapes as indicated by the dashed lines.*

To further elucidate the influence of shape on *body mass index* we show in *Figure 1* the *body mass index* of a prolate spheroidal representation of a human normalized to a sphere of the same mass (circles) or the same height (stars) in order to get a unitless *body mass index* scale. We do this for constant height (lower curve), which is an individual measure since we consider only adults and for constant mass which is a population measure (upper curve). We have indicated the typical range of aspect ratios ($\eta$) for human beings in the figure.

Consider first equal height as shown by the lower curve. The normalized *body mass index* scales as $\eta^{-2}$. The span of aspect ratios for human beings in *Figure 1* indicates a variation of almost an order of magnitude in *body mass index* ranging from severe underweight to extreme obesity, giving ample room for its use as a medical predictor. Another way to increase our understanding of the implications of this is to utilize that obviously the *body mass index* is directly proportional to the mass when keeping the height constant. We therefore get a scaling for the *body mass index* for an individual being quadratic in waistline circumference according to $BMI \propto W \propto <A>$. In this context we notice the danger for men where a broad mind and narrow waist changes places at later stages in life.

The normalized *body mass index* scales according to $\eta^{-4/3}$, where η is the aspect ratio, for individuals of the same mass. Assuming the same average body density over the population [8], and hence the same volume $BMI \propto 1/H^2 \propto (<A>)^2$. This leads to that the *body mass index* scales with the fourth power of the average waist circumference. A slight difference in (waist) circumference results in very large changes in *body mass index* and gives the requested sensitivity to be able to use it as a medical predictor in a population sense.

It is interesting to bring to mind that the resistance to (laminar) flow in arteries and veins also scales with the fourth power of their circumference. This leads to that a very small deposit of e.g. plaque or fat makes for a large impact on possible blood flow. In both cases one is led to believe that these strong fourth power dependences must have had a large impact on the evolution of living beings.

### 3. Energy aspects of the body mass index

The one and most fundamental aspect of a physical analysis after the dimensional analysis and shape study is the energy aspect. If we want to construct an index which in some sense measures the building up of unnecessary fat reserves we should compare energy in versus energy out. If the energy in is not used or lost in heat it will contribute to the building up of the reserves. For energy in to the system we use the basal metabolic rate BMR as a measure directly related to body weight as:

$$BMR = k_1 W^{3/4} \qquad (5)$$

For most animals this is a well-established scaling relationship both through extensive measurements and theoretical modelling. There is a slight variation in the exponent when one starts to go into details of various animal groups as well as intepreting data for one single group [9,10]. The prefactor is related to properties accounting for individual variations in the metabolism.

When it comes to losses we consider heat leaving the body as the major contributing source and we write down a heat loss rate (BHR) as:

$$BHR = k_2 S \qquad (6)$$

where S is the surface area giving the general scaling arguments for all individuals while the prefactor accounts for possible individual variations. The Mosteller formula for body surface area [11] being proportional to $\sqrt{WH}$ is a recommended relationship [12]. It also correlates well with later findings [13]. Using the Quetelet scaling $W \sim H^2$ we find for the ratio between basal metabolic rate and body heat loss rate:

$$\frac{energy\ stored}{energy\ lost} \propto \frac{BMR}{BHR} \propto \frac{W^{3/4}}{S} \propto \frac{W}{H^2} \equiv BMI \qquad (7)$$

Within the uncertainties in the empirical relationships used and the gross approximations made we see that our "physical" index based on the energy aspect is basically the same that Quetelet proposed almost 200 years ago based on empirical observations.

A direct physical interpretation of *body mass index* is thus to see it as the ratio of the basal metabolic rate and body heat rate loss. Since the amount of stored energy in the form of fat tissue is related to the energy turn-around, i.e. the metabolic rate and the amount of lost energy is related to surface area, we see that a large storage capacity coupled to small losses gives a large index. Since the sphere is the body with least surface area for a given volume (= energy production) we see that the index is largest for very "spherical" people. Eq.(7) is our main result in this paper and shows a direct relationship between a quick and easy number for asserting health hazards and the fundamental physical implication that if you feed in too much energy, which is not used directly, it will be stored in a more round body-shape.

## Acknowledgments

P. Apell acknowledges the invitation to J. Sabin's 70[th] anniversary conference where some of these thoughts on the *body mass index* were presented and discussed for the first time.

# References

1. R. Huxley, S. Mendis, E. Zheleznyakov, S. Reddy and J. Chan, European Journal of Clinical Nutrition **64**, *Body mass index, waist circumference and waist:hip ratio as predictors of cardiovascular risk – a review of the literature*; 16-22 (2010).

2. A. Keys, F. Fidanza, M.J. Karvonen, N. Kimura and H.L. Taylor, J. Chron. Dis. **25**, *Indices of Relative Weigth and Obesity*, 329-343 (1972).

3. A. Quetelet, *Recherches sur le poids de l'homme aux different ages.* Nouveaux Memoire de l'Academie Royale des Sciences et Belles-Lettres de Bruxelles. 1832, t. VII.

4. A. Quetelet, *A Treatise on Man and the Development of his Faculties*. Originally published in 1842. Reprinted in 1968 by Burt Franklin, New York.

5. G. Eknoyan, Nephrology Dialysis Transplantation **23**, *Adolphe Quetelet (1796-1874) – the average man and indices of obesity,* 47-51 (2008).

6. N. J. MacKay, *Scaling of human body weight with height; the Body Mass Index revisited*, J. of Biomechanics **43***, 764-66 (2009).*

7. H. Lin, *Fundamentals of zoological scaling*, American Journal of Physics 50, 72-81 (1981).

8. For an extreme being of only fat or only muscles there is a density variation of less than 20%.

9. C.R. White and R.S. Seymor, Mammalian basal metabolic rate is proportional to body mass$^{2/3}$, PNAS **100**, 4046-49 (2003).

10. G.B. West, W.H. Woodruff and J. H. Brown, *Allometric scaling of metabolic rate from molecules and mitochondria to cells and mammals,* PNAS **99**, 2473-78 (2002).

11. R.D. Mosteller, *Simplified Calculation of Body Surface Area.* N Engl J Med **317**, 1098 (1987).

12. T.K. Lam and D.T. Leung, *More on simplified calculation of body-surface area*. N Engl J Med **318**, 1030 (1988).

13. B.J.R. Bailey and G.L. Briars, *Estimating the surface area of the human body*, Statistics in medicine **15**, 1325-32 (1996).